\documentclass[conference]{IEEEtran}
\IEEEoverridecommandlockouts
\usepackage{cite}
\usepackage{amsmath,amssymb,amsfonts}
\usepackage{algorithmic}
\usepackage{graphicx}

\usepackage{textcomp}
\usepackage{hyperref}
\usepackage{multirow,booktabs}
\usepackage[table, dvipsnames]{xcolor}
\def\BibTeX{{\rm B\kern-.05em{\sc i\kern-.025em b}\kern-.08em
    T\kern-.1667em\lower.7ex\hbox{E}\kern-.125emX}}

\definecolor{LightCyan}{rgb}{0.88,1,1}
\usepackage[most]{tcolorbox}
\makeatletter



\makeatletter
\newcommand*\titleheader[1]{\gdef\@titleheader{#1}}
\AtBeginDocument{%
  \let\st@red@title\@title%
  \def\@title{%
    \bgroup\normalfont\small\raggedright\@titleheader\par\egroup
    \vskip0.1em\st@red@title}
} \makeatother

\title{Towards a Deeper Understanding of Transformer for Residential Non-intrusive Load Monitoring}
\titleheader{ }
\begin{document}

\author{\IEEEauthorblockN{Minhajur Rahman}
\IEEEauthorblockA{\textit{Dept. of Electrical \& Electronic Engineering} \\
\textit{Int'l Islamic University Chittagong}\\
Chittagong, Bangladesh \\
fahad061299@gmail.com}
\and
\IEEEauthorblockN{Yasir Arafat}
\IEEEauthorblockA{\textit{Dept. of Electrical \& Electronic Engineering} \\
\textit{Int'l Islamic University Chittagong}\\
Chittagong, Bangladesh \\
yaeeeiiuc@gmail.com}
}
\maketitle
\begin{abstract}
Transformer models have demonstrated impressive performance in Non-Intrusive Load Monitoring (NILM) applications in recent years. Despite their success, existing studies have not thoroughly examined the impact of various hyper-parameters on model performance, which is crucial for advancing high-performing transformer models. In this work, a comprehensive series of experiments have been conducted to analyze the influence of these hyper-parameters in the context of residential NILM. This study delves into the effects of the number of hidden dimensions in the attention layer, the number of attention layers, the number of attention heads, and the dropout ratio on transformer performance. Furthermore, the role of the masking ratio has explored in BERT-style transformer training, providing a detailed investigation into its impact on NILM tasks. Based on these experiments, the optimal hyper-parameters have been selected and used them to train a transformer model, which surpasses the performance of existing models. The experimental findings offer valuable insights and guidelines for optimizing transformer architectures, aiming to enhance their effectiveness and efficiency in NILM applications. It is expected that this work will serve as a foundation for future research and development of more robust and capable transformer models for NILM.
\end{abstract}


\section{Introduction}
The efficient use of energy has always been a critical challenge for humanity \cite{alahakoon2015smart}. Due to the rise in energy demand and the depletion of fossil fuel reserves, there is an ongoing global trend to adopt sustainable Renewable Energy Systems (RES). Smart grids equipped with Demand-Side Management (DSM) have the capability to modify and optimize the power consumption patterns of end users over time through skims like dynamic power pricing \cite{wilson2015smart}. With advanced RES systems, the focus shifts to smart Residential Energy Management Systems (REMS). REMS are considering data-driven approaches for modelling energy consumption behaviour because consumption patterns vary across borders. After the massive roll-out of Smart energy Meters (SMs) and Advanced Metering Infrastructures (AMIs), the NILM technology became a promising solution for modelling user energy consumption behaviour which offers 5\% to 12\% energy-saving \cite{fischer2008feedback}. 
\begin{figure}[t]
\centerline{\includegraphics[width=1\columnwidth]{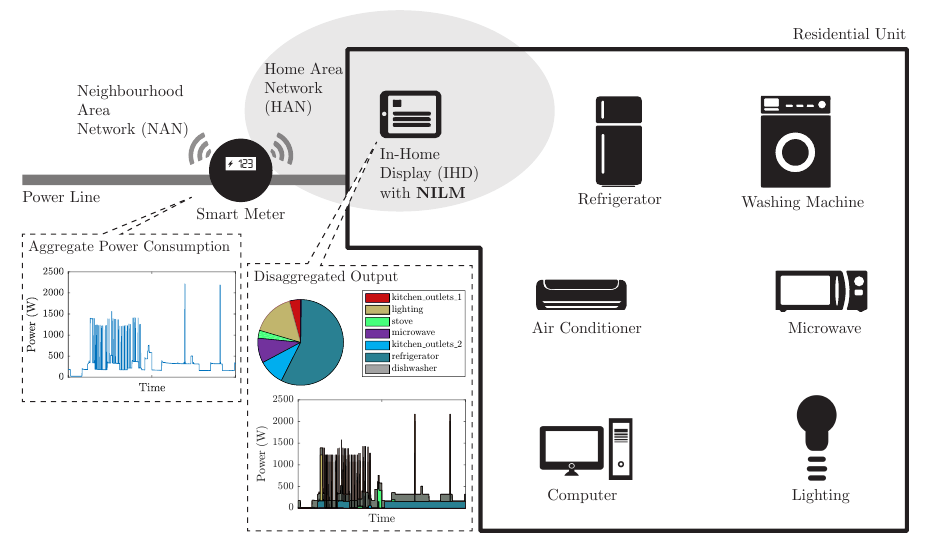}}
\caption{An overview of NILM in a residential unit. Given an aggregated load of a household, it can decompose into individual appliance-level loads for understanding energy usage and further regulation or load monitoring. }
\label{fig:NILM unit}
\end{figure}

NILM, also known as energy disaggregation, is a single-channel blind source separation problem where the aggregate level electric load is broken down into appliance-level loads in a fully automated and non-intrusive manner. This process can be viewed as the decomposition of the aggregate power signal of a household into its additive sub-components, i.e., the power signals of each domestic appliance. NILM operates at the main power entering point of the house, i.e., SMs, and communicates with the in-home display (IHD) monitor. NILM works on the basis that each appliance has an unique power drawing signature. If we can this learn uniqueness of appliance signals, NILM algorithm can access individual appliances' power usages from the total power usages of the house in a non-invasive manner. Here “non-intrusive” refers to the fact that the whole process of electric power usage monitoring and disaggregating individual’s power consumption is non-invasive and happens without any disruption of the activity or operation of the users. Figure \ref{fig:NILM unit} gives an overview of NILM in a residential setup.

NILM is a challenging task to implement in real-time. The difficulty arises from the unknown power consumption patterns of appliances. 
The presence of multiple appliances with similar power consumption patterns and additional noise signals further complicates the task. However, machine learning, particularly deep learning approaches, has made significant advancements in addressing the challenges of NILM. Below we provide a discussion of these approaches.

\subsection{Related works} 
Based on our survey, existing approaches can be divided into two mainstream approaches based on their internal mechanisms and functional algorithms, namely, \textit{classical machine learning }and \textit{deep learning} approaches. 

\textbf{Classical machine learning approaches:} These approaches assume that power signals can be broken down into a series of power features, with each feature corresponding to a different appliance. By extracting these features, the power consumption of individual appliances can be identified from the aggregated signal. Machine learning algorithms, such as support vector machine (SVM) \cite{lin2010applying}, genetic algorithm \cite{he2019efficient}, and sparse coding \cite{elhamifar2015energy}, have been used to decompose the aggregated features into the power features of each appliance. Mostly these approaches are based on event detection of the appliances. However, there are several drawbacks to these approaches. Firstly, a pre-processing step is required to extract features and tag data, which requires specialist knowledge and may introduce errors. Secondly, these approaches do not scale well to new houses or appliances, as they assume the aggregated power data is the sum of pre-defined features from training houses. In practice, the number and types of appliances in testing houses may be different. 

\textbf{Deep learning approach:} Deep learning has emerged from solving mainstream computer vision problems and influenced other data-driven research fields. Kelly and Knottenbelt are one of the first to propose a deep learning based NILM framework \cite{kelly2015neural} that outperformed classical machine learning methods. The advantage of deep learning architectures is that they can directly process the raw power signal without the need for manual feature extraction or data tagging. They also accommodate noise signals and unknown appliances without assuming that the aggregated power signal fully represents the sum of the exact individual appliance power signal.
Zhang et al. \cite{zhang2018sequence} enhanced their convolutional neural network (CNN) model's capacity which surpassed the previous methods. Deep learning research has further improved NILM algorithm's performance through various strategies, constructing hybrid models using multiple structures such as CNN, long short-term memory (LSTM) and generative adversarial network (GAN) \cite{zhou2021non, he2019generic}.
However, CNN models effectively capture the local features within power consumption sequences but they may not adequately capture global features and dependency correlations between different positions in the sequence. 

Attention mechanisms \cite{bahdanau2014neural}, originally formulated in the field of natural language processing, happen to show promising results with NILM problems, especially with transformers. Yue's BERT4NILM \cite{yue2020bert4nilm} incorporates a bidirectional transformer architecture. Followed by their work methods such as, Sykiotis's ELECTRIcity-NILM \cite{sykiotis2022electricity}, Wang's Midformer \cite{wang2022transformer}, Kamyshev's COLD,\cite{kamyshev2021cold} Yue's ELTransformer\cite{yue2022efficient}, Zhou's TTRNet\cite{zhou2022deep} employed transformer in NILM. Transformer's capability to capture long-range temporal dependencies made them highly applicable for time-series power consumption data.

\subsection{Our contribution}
While BERT4NILM pioneered the use of transformers for NILM, it provides a limited understanding of the transformer with respect to its various hyper-parameters such as the number of layers and attention heads. There are also subsequent works that improved the BERT4NILM with various training mechanisms but their provided understanding of transformer with respect to NILM is also limited. Motivated by this gap and to foster future research in NILM with transformer, in this work, a comprehensive understanding of transformer architecture for residential NILM has been provided. Specifically, comprehensive experiments on various hyper-parameters of transformer architecture such as the number of attention heads, hidden dimensions and layers have been provided. Such experiments reveal the optimal transformer model, which is crucial for obtaining better performance. Based on the survey, this work is the first to analyze the performance of transformer with such large-scale experiments. In addition, this work provides comprehensive experiments on the BERT training strategy with the masking mechanism \cite{yue2020bert4nilm} which has not been provided by the existing works. It is hope that this analysis gives researchers further understanding of transformers that are non-existent in the existing works and motivates new transformer architectures. Note that these findings can also be useful for non-residential cases.

\section{Preliminaries}
 The Transformer \cite{vaswani2017attention} is a deep learning architecture based on multi-head self-attention (MHSA) that has outperformed CNNs across various tasks. The attention mechanism \cite{vaswani2023attention} maps a query and key-value pairs from the input to produce an output representation. Queries, keys, values, and outputs are represented as matrices, and the output is computed by calculating the weighted sum of the input values, with weights based on the correlation between the query and keys.
 


\subsection{Single-head self-attention mechanism}
The single-head self-attention (scaled dot-product attention) can be formulated using the matrices \( \mathbf{Q} \) (Query), \( \mathbf{K} \) (Key), and \( \mathbf{V} \) (Value), which are obtained through linear transformations of the input matrix. The queries \( \mathbf{Q} \) and keys \( \mathbf{K} \) have dimensions \( d_k \), while the values \( \mathbf{V} \) have dimensions \( d_v \). The product of \( \mathbf{Q} \) and \( \mathbf{K}^T \) is scaled by \( \sqrt{d_k} \), then passed through a $\mathtt{softmax(\cdot)}$ operation to form soft attention, which is then multiplied by \( \mathbf{V} \) to yield a weighted value matrix. Finally, \( \mathbf{V} \) is multiplied by the correlation matrix of \( \mathbf{Q} \) and \( \mathbf{K} \) to compute the self-attention output. This process is represented by the following equation:
\begin{equation}
    \text{Attention}(\mathbf{Q}, \mathbf{K}, \mathbf{V}) = \mathtt{softmax}\left( \frac{\mathbf{Q} \mathbf{K}^T}{\sqrt{d_k}} \right) \mathbf{V}
\end{equation}
Here, the softmax function converts the matrix of real values into a probability distribution. Let \( \mathbf{X} \in \mathbb{R}^{d_k \times d_k} \) represent the matrix \( \frac{\mathbf{Q} \mathbf{K}^T}{\sqrt{d_k}} \), where \( \mathbf{X} = [X_1, X_2, \dots, X_{d_k}] \). The softmax function is applied element-wise across each \( X_i \) and can be computed as follows:
\begin{equation}\mathbf{
\mathtt{softmax}(X_{i}) = \frac{exp(X_{i,1})}{\sum_{j=0}^{d_{k}} exp(X_{i,j})}, ..., \frac{exp(X_{i,d_{k}})}{\sum_{j=0}^{d_{k}} exp(X_{i,j})}}
\end{equation}
After applying the $\mathtt{softmax(\cdot)}$ operation, the result is \( \mathtt{softmax}(\mathbf{X}) = \mathtt{softmax}\left( \frac{\mathbf{Q} \mathbf{K}^T}{\sqrt{d_k}} \right) \in [0,1]^{d_k \times d_k} \), where each row of the attention matrix sums to 1.
\subsection{Multi-head self-attention mechanism}
Multi-head attention divides the hidden space into multiple subspaces with parameter matrices and performs computations similar to the self-attention mechanism, resulting in multiple \( \mathbf{Q} \) (Query), \( \mathbf{K} \) (Key), and \( \mathbf{V} \) (Value) matrices. Each of them forms an individual attention that can access information from different subspaces. Multi-head attention linearly projects different \( \mathbf{Q}, \mathbf{K}, \mathbf{V} \) matrices \( h \) times, where \( h \) is the number of attention heads. The results are concatenated and transformed to produce the final attention output:\begin{equation} \label{MHSA_equ_2}
\begin{aligned}
MultiHead(Q, K, V) = Concat. \left(head_1, \ldots,head _h\right) W^O
\end{aligned}
\end{equation}
where, $head { }_i = Attention \left(Q W_i^Q, K W_i^K, V W_i^V\right) \text {.}$
\section{Problem Formulation}

In this section, residential NILM, which is the focus of this work, is discussed. A home consists of many appliances, which can be grouped into four categories based on their operational states:

Category I: Simple on/off appliances such as lights and microwaves. Category II: Appliances with fixed operating states and repeatable patterns, like fridges and dishwashers. Category III: Appliances with variable states, such as dimmers and power drills. Category IV: Constant-power appliances like smoke and fire detectors.
Category I and Category II appliances are the most common and consume the majority of electricity in the home. These two appliance categories are the main focus in existing work, e.g., \cite{he2019generic, yue2020bert4nilm, sykiotis2022electricity}. The formulation of the NILM problem is given as follows:

\begin{tcolorbox}
Let \( N \) be the number of appliances in a residential home, and let \( i \) be the index referring to the \( i \)-th appliance \( (i = 1, 2, \dots, N) \) \cite{kelly2015neural}. The aggregated power consumption \( \mathbf{P_{\text{agg}}}(t) \) at a time stamp \( t \) is the sum of the power consumption of individual appliances \( N \), denoted by \( \mathbf{P_i} \), where \( i = (1, 2, \dots, N) \). Thus, the NILM problem is defined as:
\begin{equation} \label{NILM_formulation}
\mathbf{P_{\text{agg}}}(t) = \sum_{i=1}^N \mathbf{P_i}(t) + \mathbf{\varepsilon_{\text{noise}}}(t)
\end{equation}
where \(\mathbf{\varepsilon_{\text{noise}}}(t) \) represents additive noise from random appliances. In a real-time NILM framework, only \( \mathbf{P_{\text{agg}}}(t) \) is observed. The task is to estimate the values \( \mathbf{\widehat{P_i}}(t) \), which are the estimated signals of each appliance, from the aggregated signal \( \mathbf{P_{\text{agg}}}(t) \).
\end{tcolorbox}
\section{Methodology}
In this section, a generalized NILM system including data acquisition, model testing, and testing steps, suitable for residential use cases has been presented. The proposed transformer model for NILM has been discussed in detail.
\begin{figure}[t]
\centerline{\includegraphics[width=1\columnwidth]{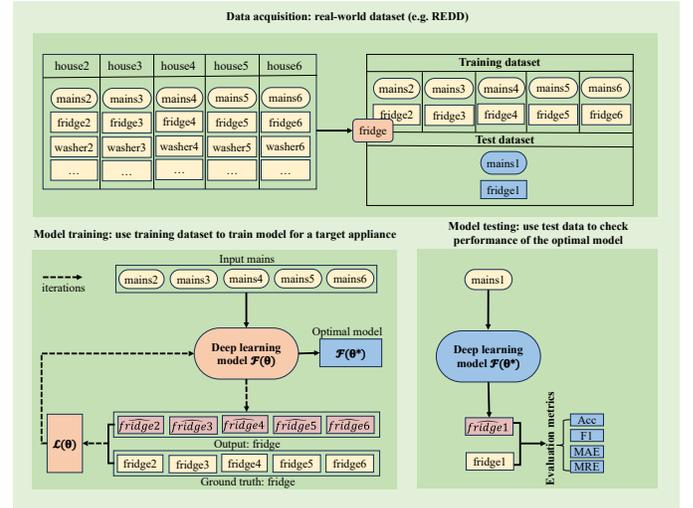}}
\caption{NILM system architecture used in this paper. Best viewed in Zoom.}
\label{fig:data_acquisition}
\end{figure}
\subsection{System Architecture}
\textbf{Data acquisition:} In a practical real-world NILM system, a large pool of power consumption data is collected from subscriber's appliances and whole home power consumption. In practice, smart meter datasets have been used for building NILM systems, e.g., REDD. These datasets provides both aggregate level consumption and individual appliance level consumption data. In this work, the REDD dataset \cite{kolter2011redd} has been used which consists of 6 house power consumption data. A particular appliance has been selected for training the model and the NILM system operates as shown in \ref{fig:data_acquisition}.  

\begin{figure*}[t]
\centerline{\includegraphics[width=1\linewidth]{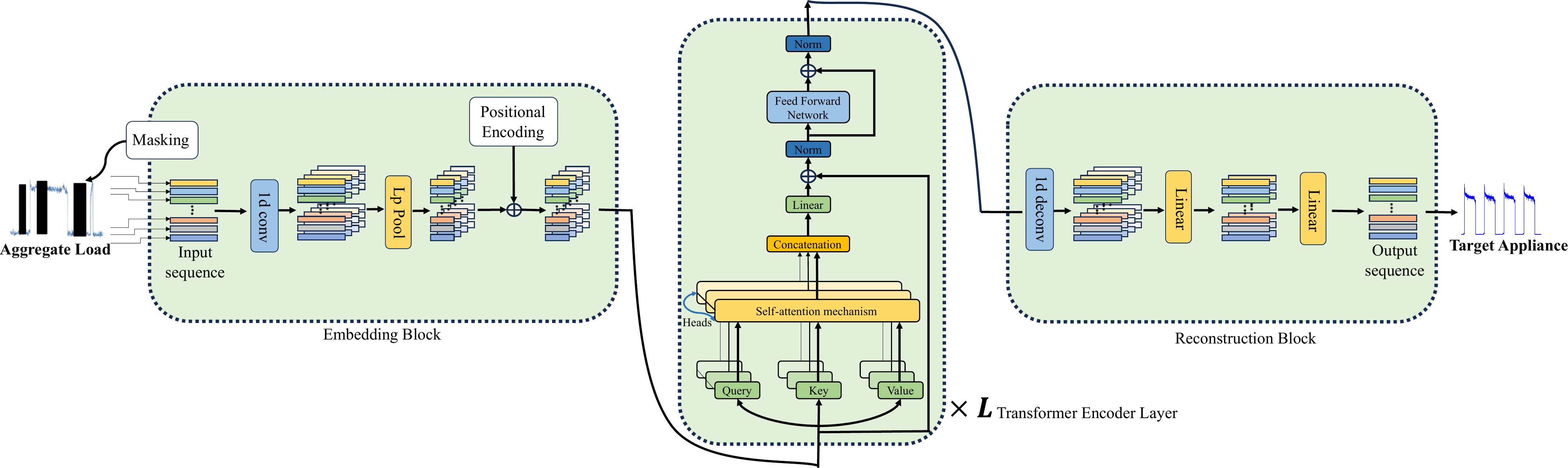}}
\caption{Transformer architecture used in this paper. Given an aggregated load, it predicts the load of a target appliance. Best viewed in Zoom.}
\label{fig:BERT model}
\end{figure*}
\textbf{Model training and testing:} If  $\mathbf{\mathcal{F(\theta)}}$ is the deep learning model and $\mathbf{\theta}$ being its corresponding parameters, the goal for model training is to make predicted appliance power sequence data ${\mathbf{(\mathcal{F}(\theta, X), Y^{i})}}$ and real appliance power data as close as possible by changing $\theta$ iteratively to minimize the loss function as follows:
\begin{equation} \mathbf{\theta} = \arg \min \mathcal{L}\left(\mathbf{\mathcal{F}(\theta, X)}, \mathbf{Y^i}\right) \end{equation}
One model $\mathcal{F(\theta)}$ is a transformer model (shown shortly) and trained for each appliance. The structure of all models are same, the difference comes from the model parameters $\theta$ (discussed shortly). The input is training data, and the model make predictions and iterates till the optimal model is selected.

The input data is aggregated channel power consumption data from the training set, e.g., $mains2$,  $mains3$, $mains4$, $mains5$, $mains6$, and the output is the predicted appliance power consumption data for selected appliance, e.g., $\widehat{fridge2}$, $\widehat{fridge3}$, $\widehat{fridge4}$, $\widehat{fridge5}$, $\widehat{fridge6})$, in the corresponding period of time. The obtained optimal model $\mathcal{F(\theta^*)}$ is tested using input $mains1$ into the obt $\mathcal{F(\theta^*)}$. As output $\widehat{fridge1}$ has been predicted. To quantify the performance of obtained model, Accuracy, F1 score, MAE (mean average error) and MRE (mean relative error) metrics has been evaluated \cite{sykiotis2022electricity} according to predicted output $\widehat{fridge1}$ and ground truth data $fridge1$. This process has been repeated for each appliance in the dataset (Fig. \ref{fig:data_acquisition} shows the training and testing process).

\subsection{Proposed transformer architecture}
This section describes the transformer model discussed above. Inspired by the BERT4NILM \cite{yue2020bert4nilm} which pre-trains a transformer with a BERT-like mechanism for energy disaggregation, we also employ a similar-looking architecture and training mechanism to study the transformers for NILM. The proposed architecture is depicted in Figure \ref{fig:BERT model}. Following the pre-processing step of the REDD dataset (the same pre-processing strategy used in \cite{yue2020bert4nilm} has been followed), input signals are passed through an embedding block to produce a set of embedding vectors. The embedding block is comprised of a convolution and pooling layer. The convolution layer $\mathbf{conv(\cdot)}$ produces a set of feature maps (which can be regarded as embedding vectors) and the pooling layer $\mathbf{pool(\cdot)}$ reduces their dimension for computational efficiency. Formally, the convolution layer computes feature maps from the input signal $X'$ as follows.
\begin{equation} \label{Conv_equ} \mathbf{
Y^i = \mathtt{conv}\left(X^{\prime}, W^i\right), W \in R^d}
\end{equation}
where $\mathbf{W}$ is learnable weight and $\mathbf{Y}$ is the feature map. Note that there can be an additional learnable bias parameter in the $\mathtt{conv(\cdot)}$. Pooling reduces the size of $\mathbf{Y}$ as follows. 
\begin{equation} \label{pooling_equ}\mathbf{
Z^i=\mathtt{pool}\left(Y^i, \alpha\right)}
\end{equation}
Positional information is found to be effective in the literature for transformers to obtain improved performance. A learnable positional vector $\mathbf{e}$ is produced and concatenate it with the pooled embedding vector $\mathbf{Z}$. Formally,
\begin{equation}\mathbf{
Z^{\prime i}=Z^i+e}
\end{equation}
where $\mathbf{Z'}$ is the final embedding vector and $\mathbf{+}$ is position-wise concatenation. Given the $\mathbf{Z'}$ transformer, blocks compute self-attention operation based on the mechanism described in the preliminaries section. The $\mathbf{Q,K,V}$ are created from $\mathbf{Z'}$ as follows. $\mathbf{Q,K,V=Z',Z',Z'}$. The output of one transformer block becomes the input of the other transformer block.
\begin{table*}[]
\caption{Results obtained with transformer architecture on REDD dataset. The results of other methods are quoted from \cite{yue2020bert4nilm}. Higher Accuracy (Acc.), F1 score (F1) and lower MRE and MAE are better.}
\resizebox{\textwidth}{!}{%
\begin{tabular}{l|l|l|cccc|cccc|cccc|cccc}
\hline\hline
 &  &  & \multicolumn{4}{c|}{\textbf{Fridge}} & \multicolumn{4}{c|}{\textbf{Washer}} & \multicolumn{4}{c|}{\textbf{Microwave}} & \multicolumn{4}{c}{\textbf{Dishwasher}} \\ \hline
 &  &  & Acc . & F1 & MRE & MAE & Acc . & F1 & MRE & MAE & Acc . & F1 & MRE & MAE & Acc . & F1 & MRE & MAE \\ \cline{2-19}
& \multirow{3}{*}{\rotatebox[origin=c]{90}{\cite{yue2020bert4nilm}}} & GRU + & 0.794 & 0.705 & 0.829 & 44.28 & 0.922 & 0.216 & 0.09 & 27.63 & 0.988 & 0.574 & 0.059 & 17.72 & 0.955 & 0.034 & 0.042 & 25.29 \\
 &  & LSTM + & 0.789 & 0.709 & 0.841 & 44.82 & 0.989 & 0.125 & 0.02 & 35.73 & 0.989 & 0.604 & 0.058 & 17.39 & 0.956 & 0.421 & 0.056 & 25.25 \\
 &  & CNN & 0.796 & 0.689 & 0.822 & 35.69 & 0.970 & 0.274 & 0.042 & 36.12 & 0.986 & 0.378 & 0.06 & 18.59 & 0.953 & 0.298 & 0.053 & 25.29 \\ \hline \hline
 
\multirow{45}{*}{\rotatebox[origin=c]{90}{Propsoed transformer architecture}} & \multirow{9}{*}{\rotatebox[origin=c]{90}{Hidden dimension}} & 4 & 0.824 & 0.738 & 0.835 & 39.03 & 0.986 & 0.611 & 0.024 & 26.93 & 0.986 & 0.435 & 0.060 & 18.55 & 0.966 & 0.520 & 0.049 & 25.08 \\
 &  & 8 & 0.871 & 0.792 & 0.836 & 35.39 & 0.988 & 0.020 & 0.000 & 35.78 & 0.987 & 0.456 & 0.058 & 17.99 & 0.964 & 0.594 & 0.055 & 23.57 \\
 &  & 16 & 0.883 & 0.808 & 0.824 & 32.38 & 0.970 & 0.419 & 0.036 & 12.92 & 0.986 & 0.495 & 0.058 & 17.92 & 0.958 & 0.497 & 0.057 & 24.18 \\
 &  & 32 & 0.881 & 0.805 & 0.812 & 28.99 & 0.988 & 0.020 & 0.000 & 35.78 & 0.985 & 0.493 & 0.061 & 18.46 & 0.966 & 0.472 & 0.042 & 21.61 \\
 &  & 64 & 0.877 & 0.799 & 0.814 & 28.86 & 0.984 & 0.142 & 0.025 & 35.18 & 0.988 & 0.453 & 0.056 & 17.09 & 0.970 & 0.525 & 0.037 & 19.90 \\
 &  & 128 & 0.855 & 0.772 & 0.811 & 30.85 & 0.982 & 0.550 & 0.031 & 28.40 & 0.988 & 0.470 & 0.057 & 17.55 & 0.966 & 0.505 & 0.044 & 21.55 \\
 &  & 256 & 0.849 & 0.765 & 0.810 & 32.90 & 0.967 & 0.493 & 0.043 & 23.26 & 0.989 & 0.547 & 0.054 & 15.62 & 0.967 & 0.493 & 0.043 & 23.26 \\
 &  & 512 & 0.750 & 0.663 & 0.810 & 43.13 & 0.966 & 0.403 & 0.040 & 12.92 & 0.988 & 0.503 & 0.057 & 17.50 & 0.962 & 0.481 & 0.051 & 24.87 \\
 &  & 1024 & 0.735 & 0.650 & 0.818 & 49.49 & 0.988 & 0.001 & 0.020 & 35.78 & 0.987 & 0.330 & 0.056 & 17.43 & 0.961 & 0.290 & 0.042 & 24.30 \\ \cline{2-19}
 & \multirow{10}{*}{\rotatebox[origin=c]{90}{No. of layers}} & 1 & 0.870 & 0.790 & 0.824 & 31.29 & 0.985 & 0.603 & 0.029 & 33.90 & 0.987 & 0.304 & 0.058 & 18.10 & 0.965 & 0.451 & 0.042 & 22.73 \\
 &  & 2 & 0.849 & 0.765 & 0.810 & 32.90 & 0.961 & 0.363 & 0.044 & 16.47 & 0.989 & 0.547 & 0.054 & 15.62 & 0.967 & 0.493 & 0.043 & 23.26 \\
 &  & 3 & 0.869 & 0.791 & 0.804 & 29.48 & 0.969 & 0.404 & 0.041 & 22.95 & 0.990 & 0.581 & 0.054 & 16.33 & 0.966 & 0.497 & 0.047 & 25.05 \\
 &  & 4 & 0.890 & 0.816 & 0.813 & 29.57 & 0.991 & 0.582 & 0.021 & 33.12 & 0.989 & 0.495 & 0.055 & 16.30 & 0.975 & 0.646 & 0.039 & 21.00 \\
 &  & 5 & 0.884 & 0.808 & 0.818 & 30.54 & 0.977 & 0.491 & 0.036 & 26.68 & 0.988 & 0.432 & 0.055 & 16.61 & 0.964 & 0.335 & 0.040 & 23.97 \\
 &  & 6 & 0.861 & 0.780 & 0.806 & 29.76 & 0.965 & 0.393 & 0.041 & 16.71 & 0.989 & 0.499 & 0.054 & 15.95 & 0.978 & 0.712 & 0.038 & 17.66 \\
 &  & 7 & 0.753 & 0.659 & 0.811 & 44.16 & 0.964 & 0.393 & 0.043 & 20.85 & 0.987 & 0.363 & 0.056 & 17.82 & 0.970 & 0.604 & 0.045 & 21.77 \\
 &  & 8 & 0.752 & 0.657 & 0.816 & 48.13 & 0.967 & 0.415 & 0.040 & 22.01 & 0.988 & 0.489 & 0.056 & 17.34 & 0.980 & 0.766 & 0.038 & 14.36 \\
 &  & 9 & 0.718 & 0.636 & 0.813 & 56.02 & 0.988 & 0.031 & 0.020 & 35.78 & 0.987 & 0.399 & 0.056 & 17.36 & 0.970 & 0.615 & 0.045 & 21.67 \\
 &  & 10 & 0.880 & 0.802 & 0.839 & 35.05 & 0.988 & 0.013 & 0.020 & 35.79 & 0.988 & 0.526 & 0.055 & 17.47 & 0.976 & 0.643 & 0.035 & 18.49 \\ \cline{2-19}
 & \multirow{8}{*}{\rotatebox[origin=c]{90}{No. of attn. heads}} & 1 & 0.840 & 0.755 & 0.805 & 32.21 & 0.977 & 0.490 & 0.029 & 16.94 & 0.987 & 0.340 & 0.057 & 18.11 & 0.964 & 0.501 & 0.046 & 19.99 \\
 &  & 2 & 0.849 & 0.765 & 0.810 & 32.90 & 0.961 & 0.363 & 0.044 & 16.47 & 0.989 & 0.547 & 0.054 & 15.62 & 0.967 & 0.493 & 0.043 & 23.26 \\
 &  & 4 & 0.886 & 0.812 & 0.808 & 28.69 & 0.992 & 0.703 & 0.019 & 24.89 & 0.988 & 0.456 & 0.057 & 17.40 & 0.965 & 0.429 & 0.042 & 23.97 \\
 &  & 8 & 0.887 & 0.814 & 0.805 & 27.90 & 0.977 & 0.492 & 0.032 & 19.80 & 0.987 & 0.446 & 0.056 & 17.07 & 0.965 & 0.405 & 0.042 & 24.65 \\
 &  & 16 & 0.889 & 0.816 & 0.806 & 28.12 & 0.988 & 0.126 & 0.020 & 35.73 & 0.987 & 0.427 & 0.056 & 16.91 & 0.965 & 0.438 & 0.044 & 24.21 \\
 &  & 32 & 0.890 & 0.817 & 0.806 & 28.16 & 0.963 & 0.383 & 0.045 & 17.71 & 0.988 & 0.479 & 0.056 & 17.31 & 0.971 & 0.537 & 0.042 & 24.59 \\
 &  & 64 & 0.888 & 0.815 & 0.803 & 27.90 & 0.991 & 0.678 & 0.023 & 34.52 & 0.987 & 0.316 & 0.057 & 18.17 & 0.961 & 0.251 & 0.042 & 25.06 \\
 &  & 128 & 0.887 & 0.812 & 0.800 & 27.26 & 0.975 & 0.467 & 0.033 & 16.16 & 0.987 & 0.354 & 0.056 & 18.40 & 0.971 & 0.538 & 0.041 & 23.92 \\ \cline{2-19}
 & \multirow{9}{*}{\rotatebox[origin=c]{90}{Dropout ratio}} & 0.1 & 0.849 & 0.765 & 0.810 & 32.90 & 0.961 & 0.363 & 0.044 & 16.47 & 0.989 & 0.547 & 0.054 & 15.62 & 0.971 & 0.556 & 0.041 & 22.86 \\
 &  & 0.2 & 0.833 & 0.747 & 0.807 & 33.78 & 0.981 & 0.540 & 0.035 & 34.74 & 0.989 & 0.514 & 0.055 & 16.68 & 0.968 & 0.502 & 0.041 & 20.98 \\
 &  & 0.3 & 0.827 & 0.741 & 0.810 & 35.90 & 0.986 & 0.622 & 0.030 & 35.24 & 0.988 & 0.441 & 0.055 & 17.00 & 0.968 & 0.467 & 0.035 & 17.64 \\
 &  & 0.4 & 0.835 & 0.750 & 0.799 & 32.18 & 0.965 & 0.391 & 0.042 & 17.83 & 0.987 & 0.452 & 0.058 & 17.89 & 0.968 & 0.518 & 0.036 & 16.85 \\
 &  & 0.5 & 0.852 & 0.770 & 0.799 & 30.93 & 0.974 & 0.455 & 0.039 & 29.07 & 0.987 & 0.468 & 0.059 & 17.96 & 0.966 & 0.477 & 0.041 & 20.40 \\
 &  & 0.6 & 0.855 & 0.773 & 0.770 & 29.49 & 0.972 & 0.449 & 0.036 & 16.25 & 0.986 & 0.455 & 0.057 & 18.12 & 0.959 & 0.554 & 0.056 & 23.96 \\
 &  & 0.7 & 0.807 & 0.717 & 0.816 & 36.84 & 0.992 & 0.663 & 0.022 & 33.93 & 0.987 & 0.454 & 0.057 & 17.65 & 0.938 & 0.368 & 0.076 & 25.49 \\
 &  & 0.8 & 0.721 & 0.640 & 0.794 & 0.79 & 0.989 & 0.670 & 0.021 & 29.21 & 0.975 & 0.493 & 0.076 & 21.64 & 0.961 & 0.475 & 0.050 & 23.60 \\
 &  & 0.9 & 0.764 & 0.644 & 0.825 & 49.59 & 0.990 & 0.682 & 0.024 & 32.52 & 0.973 & 0.476 & 0.078 & 21.33 & 0.959 & 0.477 & 0.052 & 23.89 \\ \cline{2-19}
 & \multirow{9}{*}{\rotatebox[origin=c]{90}{Masking ratio}} & 0.1 & 0.872 & 0.793 & 0.803 & 29.20 & 0.988 & 0.208 & 0.022 & 35.65 & 0.989 & 0.491 & 0.055 & 16.33 & 0.956 & 0.420 & 0.054 & 24.13 \\
 &  & 0.2 & 0.846 & 0.763 & 0.801 & 30.79 & 0.974 & 0.467 & 0.033 & 14.06 & 0.987 & 0.419 & 0.056 & 17.75 & 0.957 & 0.243 & 0.046 & 25.21 \\
 &  & 0.3 & 0.882 & 0.806 & 0.806 & 28.02 & 0.961 & 0.373 & 0.045 & 20.63 & 0.988 & 0.482 & 0.055 & 17.00 & 0.967 & 0.523 & 0.044 & 21.90 \\
 &  & 0.4 & 0.874 & 0.795 & 0.804 & 28.74 & 0.967 & 0.390 & 0.041 & 18.02 & 0.988 & 0.439 & 0.057 & 18.05 & 0.961 & 0.374 & 0.046 & 25.63 \\
 &  & 0.5 & 0.846 & 0.761 & 0.819 & 33.16 & 0.963 & 0.377 & 0.050 & 32.43 & 0.989 & 0.530 & 0.056 & 17.19 & 0.938 & 0.112 & 0.062 & 26.49 \\
 &  & 0.6 & 0.735 & 0.647 & 0.814 & 40.06 & 0.975 & 0.430 & 0.036 & 26.71 & 0.988 & 0.470 & 0.058 & 17.93 & 0.951 & 0.112 & 0.048 & 25.33 \\
 &  & 0.7 & 0.857 & 0.774 & 0.823 & 34.26 & 0.966 & 0.399 & 0.044 & 25.92 & 0.987 & 0.475 & 0.059 & 18.14 & 0.955 & 0.0 & 0.041 & 25.28 \\
 &  & 0.8 & 0.869 & 0.788 & 0.836 & 34.49 & 0.972 & 0.399 & 0.041 & 36.58 & 0.988 & 0.516 & 0.058 & 17.52 & 0.955 & 0.000 & 0.041 & 25.28 \\
 &  & 0.9 & 0.869 & 0.789 & 0.836 & 34.43 & 0.972 & 0.399 & 0.041 & 36.58 & 0.988 & 0.516 & 0.058 & 17.52 & 0.955 & 0.000 & 0.041 & 25.28 \\
 \hline \hline
\multicolumn{3}{c|}{{BERT4NILM \cite{yue2020bert4nilm}}} & 0.841	& 0.756	& 0.806	& 32.35	& 0.991	& 0.559	& 0.022	& 34.96	& 0.989	& 0.476	& 0.057	& 17.58	& 0.969	& 0.523	& 0.039	& 20.49 \\
\rowcolor{lightgray}
\multicolumn{3}{c|}{\textbf{Proposed model }} & 0.848	& 0.765	& 0.830	& 31.16	& 0.988	& 0.633	& 0.021	& 27.47	& 0.987	& 0.496	& 0.049	& 16.30	& 0.961	& 0.542	& 0.038	& 19.03  \\
\hline \hline
\end{tabular}}
\label{tab: results}
\end{table*}

Given the output of the transformer block $F$, the reconstruction blocks perform the reconstruction of the original signal. It has two steps, namely, deconvolution and linear projection. In the deconvolution step, we perform a deconvolution operation to recover the size of the feature maps before the pooling operation. The linear projection operation reduces the feature maps to the original size. Formally, the deconvolution operation performs the following.
\begin{equation}
    \mathbf{M^i = \mathtt{tanh}(\mathtt{deconv}(F, W^i)), W \in R^d\ and\ M \in R^{n\times d}}
\end{equation}
where $\mathbf{tanh}$ is an activation function such as ReLU. The linear projection operation performs the following operation.
\begin{equation}\mathbf{
    Q = \mathtt{linear}(M, W^i), W \in R^d\ and\ Q \in R^{d}}
\end{equation}
For training stability, layer normalization and dropout operations has been performed during training. Various combinations of transformer blocks has been used which will be discussed in the following section.

\textbf{Loss function:} The below loss function is used for training. 
\begin{equation} \label{loss_function}
\begin{gathered}
\mathcal{L}(\mathbf{x}, \mathbf{s}) = \frac{1}{T} \sum_{i=1}^T (\hat{\mathbf{x}} - \mathbf{x})^2 + \frac{1}{T} \sum_{i=1}^T \log \left(1 + \exp \left(-\hat{\mathbf{s}} \mathbf{s}_i\right)\right) + \\
D_{KL}\left(\mathbf{softmax}\left(\frac{\hat{\mathbf{x}}}{\tau}\right) \middle\| \mathbf{softmax}\left(\frac{\mathbf{x}}{\tau}\right)\right) 
+ \frac{\lambda}{T} \sum_{i=0} \left|\hat{\mathbf{x}}_i - \mathbf{x}_i\right|
\end{gathered}
\end{equation}
where $\hat{\mathbf{x}}, \mathbf{x} \in [0,1]$ represent the ground truth and prediction of the output sequence divided by the maximum power limit. $\hat{\mathbf{s}}, \mathbf{s} \in \{-1,1\}$ are the appliance state label and prediction. $T$ stands for the total time steps, i.e., the sequence length, and $\mathbf{O}$ refers to the set of time steps when either the status label is on or the prediction is incorrect. In this equation, we also introduce hyperparameters $\mathbf{\tau}$ and $\lambda$ for tuning the sequence softmax temperature and reduction of absolute error.

\section{Experimental Settings and Results}
\subsection{Experimental settings and baselines}
Experiments with the REDD dataset has been conducted following the data pre-preprocessing, training and testing protocols used in \cite{yue2020bert4nilm}. There are six house data in the REDD dataset from which house 1 data for testing and the remaining house data for training has been used. Pytorch has been used to develop the transformer and conduct the experiments. The models are trained from 100 epochs with AdamW optimiser using a P100 GPU on a cloud HPC. The obtained results has been compared with GRU+, LSTM+ and CNN \cite{yue2020bert4nilm}.

Experiment with the following hyper-parameters of transformers, namely, hidden dimension of attention layers, number of transformer layers, number of attention heads and dropout ratio in attention layers has been done. For a comprehensive understanding, experiment with a large variety of these hyper-parameters have been conducted. Table \ref{tab: results} provides the list of hyper-parameters and their different values. Since the combinations of these hyper-parameters are large in number, only experiment with one hyper-parameter at a time has been conducted while keeping the other hyper-parameters fixed. The optimal hyper-parameter setting proposed in \cite{yue2020bert4nilm}, i.e., hidden dimension = 256, number of layer/attention heads = 2 and dropout ratio = 0.1 has been used for fixing the hyper-parameters that are not being investigated. In addition, experiment with the masking ratio of BERT training has been conducted. In \cite{yue2020bert4nilm}, masking ratio of 0.25 is used.

\begin{figure}[t]
\centerline{\includegraphics[width=1\columnwidth]{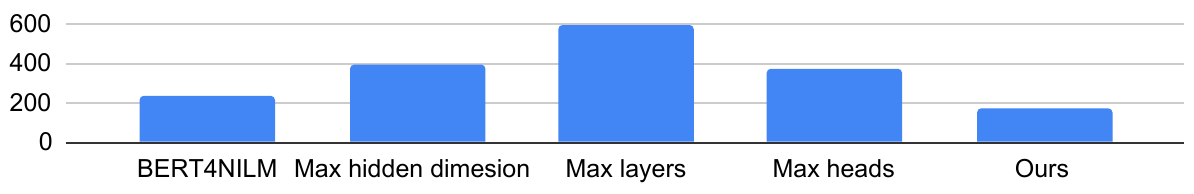}}
\caption{Computation time/epoch (in secs). Max. value in Table \ref{tab: results} is compared.}
\label{fig: time}
\end{figure}

\subsection{Analysis of results}

\subsubsection{Evaluation of various number of layers} The number of layers plays a vital role in defining the capacity of the models, i.e., the number of learnable parameters. As shown in Table \ref{tab: results}, higher capacity models yield better MAE and MRE metrics, however, at the expense of learnable parameters. We noticed that two layers are sufficient for obtaining good results across all four metrics. The higher capacity models usually require large samples to learn effective features which are non-existent in REDD dataset.

\subsubsection{Evaluation of various numbers of attention heads} Attention heads are also related to the model capacity. Based on the Table \ref{tab: results} results, 4, 32 and 128 heads produce better results, however, they contribute to higher computational cost. In comparison, two heads produce comparable results at a lower cost, therefore, we suggest using only two heads. 

\subsubsection{Evaluation of size of hidden dimension} Hidden dimension size is crucial for defining model capacity to capture complex patterns and relationships in data. Higher hidden dimension size contributes to larger learnable parameters and higher computation costs. Based on the Table \ref{tab: results} results, 16 appears to be an optimal hidden dimension size overall while higher sizes sometimes bring extra performance for some appliances.

\subsubsection{Evaluation of dropout ratio and masking ratio} Dropout plays the role of regularisation to prevent issues such as overfitting. Based on the Table \ref{tab: results} results, we think that a dropout ratio of 0.5 appears to be best among other ratios. Masking ratio selection for BERT training is also essential for good performance. Based on the obtained results, 0.3 appears to be the most effective masking ratio.

\section{Optimal model selection}
Based on the analysis above, the transformer model has been selected with the optimal hyper-parameter. Table \ref{tab: results} (last row) shows the results. It is clear that the optimal transformer has surpassed the original BERT4NILM architecture across various metrics on all four appliances. This finding justifies that selecting optimal hyper-parameters is crucial for obtaining good performance with transformer architecture. As shown in Fig. \ref{fig: time}, our optimal transformer trains more efficiently than other ones with higher capacity. 

\section{Conclusion and future works}
In this paper, a comprehensive analysis of transformer model have been performed. Based on the findings, it is clear that the transformer is robust to a large set of model hyper-parameters. Based on this observation, a compact transformer model has been proposed that uses 150x less number of parameters than the BERT4NILM transformer. The proposed compact transformer can achieve good performance on four appliances of REDD datasets. In the future, experiments can be conducted with larger datasets. Moreover qualitative analysis of results can be conducted.


\bibliographystyle{IEEEtran}
\bibliography{M.Rahman_Y.Arafat}
\end{document}